\documentclass[10pt,letterpaper]{article}

\usepackage[margin=1in]{geometry}
\usepackage[font=small,labelfont=bf,labelsep=period]{caption}

\usepackage{newtxtext,newtxmath}
\usepackage{setspace}
\usepackage{eso-pic}
\usepackage{abstract}

\usepackage{amssymb}
\usepackage{cite}  
\usepackage{enumitem, graphicx, amsmath, amssymb, sectsty, titlesec, booktabs, float, siunitx, subfigure, soul, xcolor, multirow, pbox}
\usepackage{float}
\usepackage{amsmath, amssymb}
\usepackage[activate={true,nocompatibility},final,tracking=true,kerning=true,spacing=true,factor=1100,stretch=10,shrink=10]{microtype}

\usepackage{hyperref}
\hypersetup{
  colorlinks = true, 
  linkcolor = blue, 
  anchorcolor = blue,
  citecolor = blue,  
  filecolor = blue, 
  urlcolor = blue
}

\usepackage[capitalise,nameinlink]{cleveref}
\usepackage[numbers,sort&compress]{natbib}
\usepackage[justification=centering]{caption}

\usepackage{graphicx}
\graphicspath{{files/}}
\sectionfont{\fontsize{12}{15}\selectfont}
\subsectionfont{\fontsize{12}{15}\selectfont}
\newcommand{\macaulay} [1] {\left\langle #1 \right\rangle}  

\makeatletter
\def\@maketitle{%
	\begin{center}%
		\textbf{\fontsize{14}{17}\selectfont \@title}
	\end{center}%
}
\makeatother

\usepackage{fancyhdr}

\makeatletter
\def\@maketitle{%
  \begin{center}%
    {\fontsize{16}{20}\selectfont \@title}\\[1ex]
  \end{center}%
}
\makeatother

\begin{document}

\AddToShipoutPicture*{
  \put(50,25){\parbox[b]{\textwidth}{
    \footnotesize
    \hrulefill \\
    \textbf{*Corresponding author:} Md Mahmudul Hasan Pathik, 
    \textit{Email:} \texttt{mhpathik@ncsu.edu}, \textit{Tel:} 919-438-6339
  }}
}

\title{On the Mathematical Foundation of a Decoupled Directional Distortional Hardening Model for Metal Plasticity in the Framework of Rational Thermodynamics}

\maketitle

\begin{center}
    \text{Md Mahmudur Rahman$^{a}$, Md Mahmudul Hasan Pathik$^{b,*}$, Nazrul Islam$^{c}$}
\end{center}

\begin{flushleft}
\begin{center}\small
\textit{$^{a}$School of Mechanical Engineering, Ray W. Herrick Laboratories, Purdue University, IN 47907, USA\\[0.10\baselineskip]
     $^{b}$Mechanics and Materials Lab, North Carolina State University, NC 27606 USA\\[0.10\baselineskip]
     $^{c}$ Department of Civil Engineering, Bangladesh University of Engineering and Technology, Dhaka-1000}

\end{center}
\end{flushleft}

\setstretch{1.2}

\vspace{0.5em}
\begin{center}
  {\itshape Dedicated to Jacob Lubliner}
\end{center}
\vspace{0.5em}

    \begin{abstract}
\noindent
This study proposes a modification to the yield condition that addresses the mathematical constraints inherent in the Directional Distortional Hardening models developed by Feigenbaum and Dafalias. The modified model resolves both the mathematical inconsistency found in the ‘complete model' and the limitations of the ‘\textit{r}-model'. In the complete model, inconsistency arises between the distortional term in the yield surface and the plastic part of the free energy in the absence of kinematic hardening. Additionally, the ‘\textit{r}-model' fails to capture the flattening of the yield surface in the reverse loading direction due to the absence of a fourth-order anisotropic tensor structure in the distortional term. To address these issues, the proposed model introduces a decoupled distortional hardening term in the yield function. This modification enables the simultaneous representation of both flattening and sharpening of the yield surface, and permits isotropic hardening with distortion even without kinematic hardening. A consistent mathematical formulation based on rational thermodynamics and a corresponding numerical algorithm are also developed, establishing a foundation for future experimental investigations and model validation.

\vspace{0.5em}
\noindent
Keywords:
Directional Distortional Hardening, Yield Surface, Isotropic Hardening, Kinematic Hardening.
\end{abstract}

\section{Introduction}

\begin{figure}[H]
    \centering
    \includegraphics[width=0.45\linewidth, height=6cm]{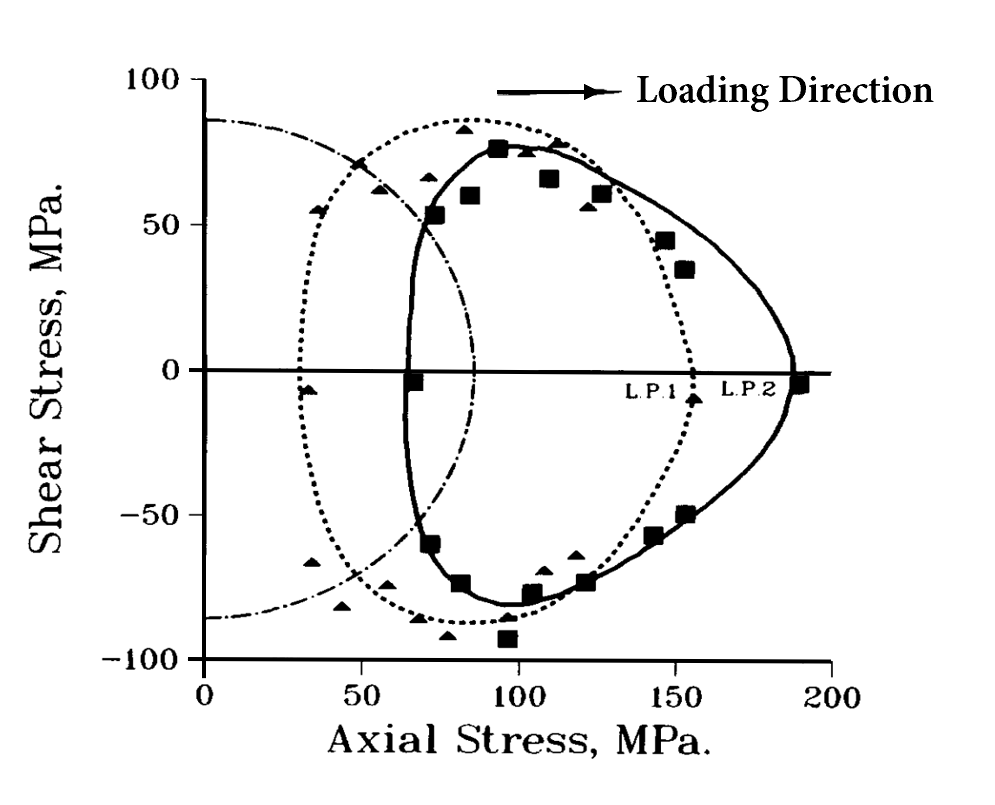}
    \caption{Experiments by Wu and Yeh \cite{wu1991experimental} on Stainless Steel
304 tubular specimens prestrained axially to loading points (L.P.) 1 and 2 in two-dimensional stress space}
    \label{fig:fg1}
\end{figure}

Directional Distortional Hardening (DDH) is a consequence of the induced anisotropy exhibited by some metals under plastic deformation and may appear in combination with other modes of hardening such as kinematic and isotropic hardening. It can be thought of as a deformation based concept in material behavior that represents the transformation of the yield surface, characterized by the development of a region with high curvature in the loading direction and flattening in the reverse direction of loading (see \cref{fig:fg1}). The directional asymmetry in yield surfaces has been well-captured in early experimental studies. Wu and Yeh \cite{wu1991experimental}, Phillips et al. \cite{phillips1974some}, McComb \cite{mccomb1960some}, Naghdi et al. \cite{naghdi1958experimental} , and Boucher et al. \cite{boucher1995experimental} were able to capture this phenomenon in the experiments. These observations provide strong motivation for constitutive models that incorporate distortional hardening through evolving anisotropic internal variables.
Various attempts have been made over the years to mathematically model the yield surface to explain the phenomenon of Directional Distortional Hardening. One of the most notable works is by Baltov and Sawczuk \cite{baltov1965rule} where the author introduced a fourth-order anisotropic tensor in the functional representation of yield surface. This fourth-order anisotropic tensor is the internal variable used to model the distortion. Dafalias et al. \cite{dafalias2003simple} proposed a yield surface model, where anisotropy is developed through kinematic and distortional hardening and distortion is represented by a fourth-order tensor $\mathbf{H}$. In this model, the evolution equation of the internal variables was developed using the principles of thermodynamics. Voyiadjis and Foroozesh \cite{10.1115/1.2897056} used the same yield surface as Dafalias et al. \cite{dafalias2003simple}. However, the fourth-order distortional tensor $\mathbf{H}$ is defined differently. Voyiadjis and Foroozesh \cite{10.1115/1.2897056} used kinematic considerations to develop the evolution equation for this fourth-order tensor unlike the work of Dafalias et al. \cite{dafalias2003simple} where thermodynamic considerations were used in this purpose.
Other attempts to model directional distortion have been made by Ortiz and Popov \cite{ortiz1983distortional}, Kurtyka and {\.Z}yczkowski \cite{kurtyka1996evolution} and Fran\c{c}ois \cite{franccois2001plasticity}. Ortiz and Popov  \cite{ortiz1983distortional} used a Fourier expansion of the harmonic function to model the yield surface. Kurtyka and Zyczkowski \cite{kurtyka1996evolution} used five hyperspheres with different radius and centers to model the distortion of the yield surface. Fran\c{c}ois \cite{franccois2001plasticity} included the directional distortion term within the context of thermodynamics for the first time. Unlike the models relying on fourth-order internal variables, his approach introduced distortion by redefining the stress measure itself using a distorted stress formulation governed by second-order ranked internal variables to describe yield surface evolution under non-proportional loading. Feigenbaum and Dafalias \cite{feigenbaum2007directional} developed a ‘complete model’ based on the work of Dafalias et al. \cite{dafalias2003simple}. In the ‘complete model’ \cite{feigenbaum2007directional}, Feigenbaum incorporated the directionality of the distortion into the yield surface equation. The equations used in this model are derived based on the second law of thermodynamics, specifically the dissipation inequality. This model suggests that energy is released during directional distortion, stored during kinematic and isotropic hardening, and uses evanescent memory type hardening rules. However, the model relies on the back-stress tensor to determine the directionality of distortion, and hence when back-stress equals zero, the model reduces to an isotropic von Mises type model, even if the anisotropy tensor $\mathbf{A}$ is non-zero. And the ‘\textit{r}-model’ \cite{feigenbaum2008simple} is also developed subsequently by Feigenbaum and Dafalias \cite{feigenbaum2008simple} which introduced an orientational second-order tensor ($\mathbf{r}$) instead of the back-stress ($\boldsymbol{\alpha}$) for the scalar multiplier. The ‘\textit{r}-model'
replaces the fourth-order distortion tensor $\mathbf{H}$ with the scalar quantity $(1 - \mathbf{n^r}:\mathbf{r})$, where $\mathbf{n^r}$ is the unit radial tensor from the center of the yield surface to the current stress state 
and  $(\mathbf{n^r}:\mathbf{r})$ is completely responsible for directional distortion. Therefore, it becomes possible to decouple kinematic hardening from distortional hardening in this model. The mathematical expression for \( \mathbf{n}^r = \frac{\mathbf{s} - \boldsymbol{\alpha}}{\lVert \mathbf{s} - \boldsymbol{\alpha} \rVert} \), where \( \mathbf{s} \) is the deviatoric part of the stress tensor and \( \boldsymbol{\alpha} \) is the backstress tensor. Due to the absence of the fourth-order anisotropic term $\mathbf{A}$ in the equation of the yield surface, the flattening of the yield surface in the opposite direction of loading cannot be captured using this model. 

Several other notable studies have investigated distortional hardening theory, emphasizing on computational efficiency and simpler internal variable structures. The distortional hardening was captured with the help of second-rank internal variables \cite{shutov2012viscoplasticity,panhans2006viscoplastic} demonstrating its suitability to closely capture the shape of the yield surfaces from experiments. Shutov et al. \cite{shutov2011phenomenological} developed a finite-strain viscoplasticity model that incorporates distortional hardening, along with isotropic and kinematic hardening, using a second-rank tensor to represent the evolution of anisotropy in the yield surface. Shutov and Ihlemann\cite{shutov2012viscoplasticity} further extended these ideas by demonstrating that arbitrary convex yield surfaces can be interpolated using evolving second-rank tensors, facilitating smooth and thermodynamically consistent transitions between experimentally observed yield shapes. Later, Marek \cite{marek2019numerical} investigated numerical implementations of distortional hardening models and briefly proposed a modification that is conceptually similar to the one developed in this work, aiming to distort the yield surface in a simplified and computationally efficient manner. Although the distortion mechanism he outlined, particularly the anisotropic term $\mathbf{A}$= $\mathbf{z}$$ \otimes $$\mathbf{z}$, where $\mathbf{z}$ represents a second-rank deviatoric tensor that defines the main axis of the distorted yield surface, was not further pursued in the dissertation, it exhibits similar asymmetry in the yield surface derived in this study.

The motivation of this research stems from the need to address specific mathematical inconsistencies within the ‘complete model' \cite{feigenbaum2007directional} and limitations in the ‘\textit{r}-model' \cite{feigenbaum2008simple} proposed by Feigenbaum and Dafalias. Specifically, our investigation focuses on rectifying issues that arise when kinematic hardening is absent in the ‘complete model' and the challenge of accurately capturing the flattening of the yield surface in the ‘\textit{r}-model' \cite{feigenbaum2008simple}. The primary contribution of this paper lies in the introduction of a modified yield surface model, followed by the mathematical formalism of the associative hardening rules, which effectively resolves the issues associated with the complete and ‘\textit{r}-model'.
\section{Framework of Feigenbaum and Dafalias Distortional Hardening Model}
\label{sec:2}

\begin{figure}[H]
    \centering
    \includegraphics[width=0.40\linewidth]{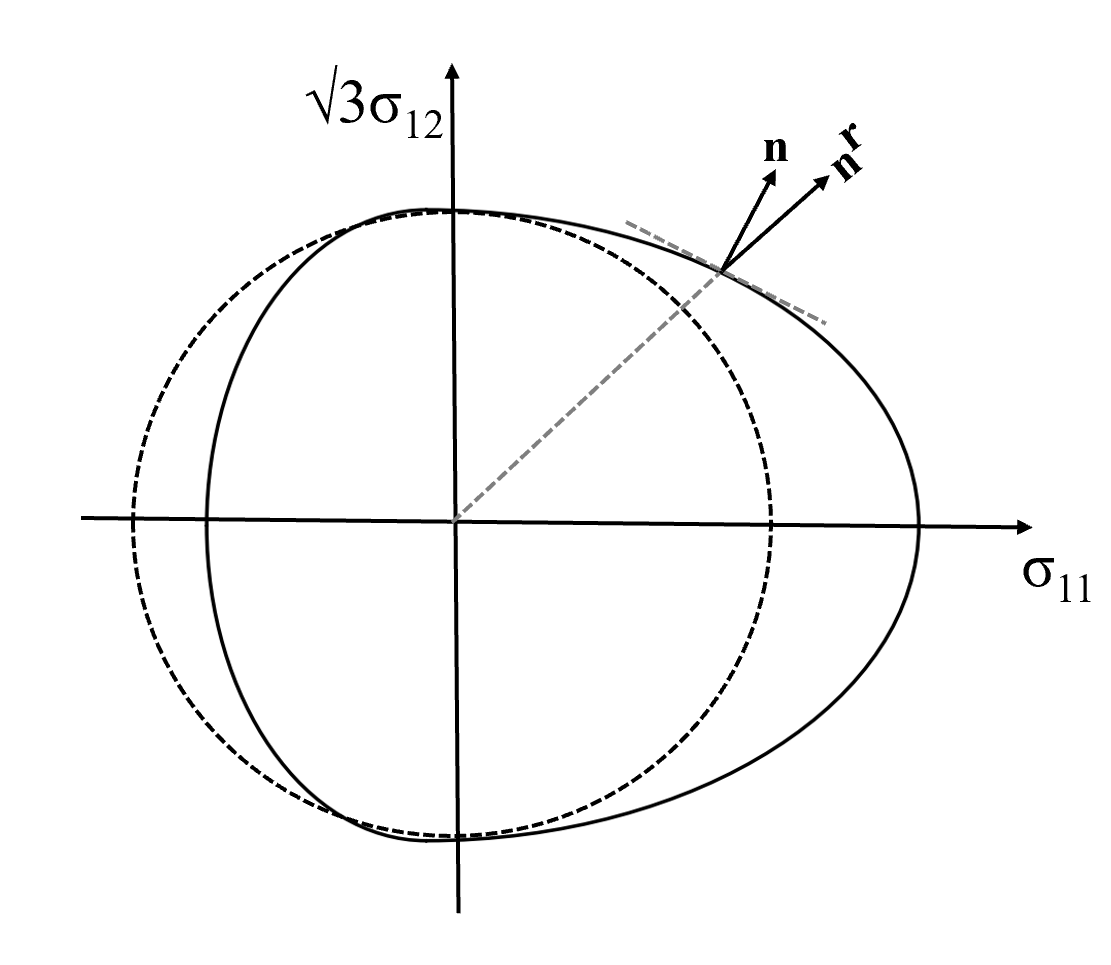}
    \caption{Schematic of the unit radial tensor ($\mathbf{n}^{r}$) and unit normal ($\mathbf{n}$) to the subsequent yield surface}
    \label{fig:fg2}
\end{figure}

Feigenbaum and Dafalias \cite{feigenbaum2007directional} developed a distortional hardening model called the ‘complete model’, which is based on the assumption of small strain. The additive strain decomposition considered for this model is shown in Eq.~(\ref{eq:strain}).

\begin{equation}
\boldsymbol{\varepsilon} = \boldsymbol{\varepsilon}_e + \boldsymbol{\varepsilon}_p.
\label{eq:strain}
\end{equation}
In this model, a modified form of the von Mises yield surface, shown in Eq.~(\ref{eq:yield_surface}), is adopted to accommodate the distortion and translation of the yield surface.

\begin{equation}
f = (\mathbf{s} - \boldsymbol{\alpha}):\{\mathbf{H}_0 + (\mathbf{n^r}:\boldsymbol{\alpha})\mathbf{A}\}:(\mathbf{s} - \boldsymbol{\alpha}) - k^2 = 0.
\label{eq:yield_surface}
\end{equation}
In this model, three internal variables $\mathbf{A}$, $\boldsymbol{\alpha}$, and $k$ are used to capture the hardening phenomenon of the metals. Here, $\mathbf{s}$ is the deviatoric stress tensor, $\boldsymbol{\alpha}$ represents the deviatoric back-stress tensor, and $k$ is the isotropic hardening parameter representing the size of the yield surface. $\mathbf{A}$ in Eq.~(\ref{eq:yield_surface}) is the fourth-order anisotropic tensor. $\mathbf{H}_0$ is $3/2$ times the deviatoric projection tensor of rank 4, and $\mathbf{n^r}$ is the unit radial tensor as shown in \cref{fig:fg2}.
The expressions of $\mathbf{H}_0$ and $\mathbf{n^r}$ are given in Eqs.~(\ref{eq:H0ijkl}) and (\ref{eq:nprime}), respectively.

\begin{equation}
H_{0ijkl} = \frac{3}{2} I_{ijkl} = \frac{3}{2} \left[ \frac{1}{2} \left( \delta_{ik}\delta_{jl} + \delta_{il}\delta_{jk} \right) - \frac{1}{3} \delta_{ij}\delta_{kl} \right]
\label{eq:H0ijkl},
\end{equation}

\begin{equation}
\mathbf{n^r} = \frac{\mathbf{s} - \boldsymbol{\alpha}}{\|\mathbf{s} - \boldsymbol{\alpha}\|}
\label{eq:nprime}.
\end{equation}

$\mathbf{H}_0$ has a property such that if this tensor is contracted with another tensor of second rank, it gives the deviatoric part of that respective tensor times $3/2$.
\noindent The model considers the associative plastic flow rule as shown in Eq.~(\ref{eq:flow_rule}):
\begin{equation}
\dot{\boldsymbol{\varepsilon}}_p = \lambda \frac{\partial f}{\partial \boldsymbol{\sigma}}
\label{eq:flow_rule}.
\end{equation}
Here, $\lambda$ is the plastic multiplier or loading index. The expression for $\lambda$ as shown in \cref{eq:lambda} can be derived using \cref{eq:flow_rule} and the consistency condition for yield function.
\begin{equation}
\lambda = \left< \frac{1}{K_p} \frac{\partial f}{\partial \boldsymbol{\sigma}} : \dot{\boldsymbol{\sigma}} \right>
\label{eq:lambda}.
\end{equation}
$K_p$ is the plastic modulus, which can be determined from the consistency condition. Here, $\frac{\partial f}{\partial \boldsymbol{\sigma}}$ is the gradient of the yield function.
The Macaulay brackets $\macaulay{}$ are used such that, the loading index $\lambda$ only has an effect when the quantity inside the brackets is positive; otherwise, the bracketed term evaluates to zero regardless of $\lambda$'s value. They serve as a mathematical tool to categorize different loading/unloading conditions based on the sign of the loading index $\lambda$, where $\lambda$$>$$0$ for plastic loading, and $\lambda$=$0$ for neutral loading case. 

\noindent Evolution equations of isotropic, tensor-valued internal variables, and anisotropic tensors are shown in Eqs.~(\ref{eq:k_dot})--(\ref{eq:A_dot}), respectively, which are chosen in a way that, they satisfy the Clausius-Duhem inequality:

\begin{equation}
\dot{k} = \lambda k\kappa_1\left( 1 - \kappa_2 k \right),
\label{eq:k_dot}
\end{equation}

\begin{equation}
\dot{\boldsymbol{\alpha}} = \lambda \left\| \frac{\partial f}{\partial \boldsymbol{\sigma}} \right\| a_1 \left( \mathbf{n} - a_2 \boldsymbol{\alpha} \right),
\label{eq:alpha_dot}
\end{equation}

\begin{equation}
\dot{\mathbf{A}} = -\lambda A_1 \left\| \mathbf{s} - \boldsymbol{\alpha} \right\|^2
 \big[ \left( \mathbf{n}^r :  \boldsymbol{\alpha} \right) \mathbf{n}^r \otimes \mathbf{n}^r + \frac{3}{2} A_2 \mathbf{A} \big].
\label{eq:A_dot}
\end{equation}
Here, $\kappa_1$, $a_1$, $A_1$ and $\kappa_2$, $a_2$, and $A_2$ are the material constants, the symbol "$\otimes$" means dyadic product or tensor product and $\mathbf{n}$ is the unit normal to the yield surface. The ‘complete model' has some inconsistencies, which are discussed in the next section.

\subsection{Complete Model and Its Inconsistency}
\label{sec:intro}

The ‘complete model' \cite{feigenbaum2007directional} is based on Coleman-Noll procedure-type rational thermodynamics~\cite{hutter2018coleman}  and includes a fourth-order tensor-valued internal variable ($\mathbf{A}$) that describes the evolving directional distortion. Hardening equations for the internal variables, as given in Eqs.~(\ref{eq:k_dot}) to (\ref{eq:A_dot}), are derived based on sufficient conditions to satisfy the thermodynamic requirements of positive energy dissipation. Evolution equtions of the internal variables are  all fading memory-type equations. 

As the associative flow rule is used in this model and plastic strain rate depends on the gradient of the yield function, the developed ‘complete model' \cite{feigenbaum2007directional} attempts to represent the yield surface with a high degree of accuracy to properly predict the directional distortional hardening.

In this model, the anisotropic term ($\mathbf{A}$) appears in a coupled form with the back-stress ($\boldsymbol{\alpha}$) in the fourth-order distortional term ($\mathbf{H}$), where $\mathbf{H} = \mathbf{H}_0 + (\mathbf{n^r} : \boldsymbol{\alpha}) \mathbf{A}$, as given in Eq.~(\ref{eq:yield_surface}). As a result, if the back-stress ($\boldsymbol{\alpha}$) is zero, distortion ($\mathbf{H}$) ceases to exist or becomes constant, and the yield surface reduces to the von Mises yield surface . Furthermore, the evolution equation of the two internal variables, $\boldsymbol{\alpha}$ and $\mathbf{A}$, calculated from the second law of thermodynamics, suggests that these can exist simultaneously even though one of these terms becomes zero, which is evident from Eqs.~(\ref{eq:k_dot})--(\ref{eq:A_dot}). 

In Eq.~(\ref{eq:psi_p}), the plastic part of the free energy per unit mass ($\psi_p$) is decomposed into isotropic and anisotropic parts, as postulated by Feigenbaum and Dafalias \cite{feigenbaum2007directional}:

\begin{equation}
\psi_p = \psi_p^{\text{iso}} + \psi_p^{\text{ani}}.
\label{eq:psi_p}
\end{equation}

The anisotropic part of the plastic free energy is further additively decomposed into two parts, which correspond to kinematic and distortional hardening. Feigenbaum and Dafalias \cite{feigenbaum2007directional} postulated that the kinematic hardening corresponds to energy storage, and the distortional hardening corresponds to energy release. Hence, the expression of the anisotropic plastic free energy follows Eq.~(\ref{eq:psi_ani}), and the total plastic energy follows Eq.~(\ref{eq:psi_total}):

\begin{equation}
\psi_p^{\text{ani}} = \psi_p^{\text{kin}} - \psi_p^{\text{dis}},
\label{eq:psi_ani}
\end{equation}

\begin{equation}
\psi_p = \psi_p^{\text{iso}} + \psi_p^{\text{kin}} - \psi_p^{\text{dis}}.
\label{eq:psi_total}
\end{equation}

The isotropic, kinematic, and distortional parts of the free energy are positive definite, and their evolution equations asymptotically approach a saturation limit. In addition, if the internal variable of a hardening term is zero, the corresponding free energy also becomes zero. For example, if $\boldsymbol{\alpha} = \boldsymbol{0}$, $\psi_p^{\text{kin}} = \frac{1}{2 a_1} \boldsymbol{\alpha} : \boldsymbol{\alpha} =  0$, and Eq.~(\ref{eq:psi_total})reduces to Eq.~(\ref{eq:psi_reduced}):

\begin{equation}
\psi_p = \psi_p^{\text{iso}} - \psi_p^{\text{dis}}.
\label{eq:psi_reduced}
\end{equation}

In the absence of kinematic hardening, the anisotropic term ($\mathbf{A}$) evolves, as evident from Eq.~(\ref{eq:A_dot}). The distortional part of the free energy $\psi_p^{\text{dis}}$ exists and can be expressed as shown in Eq.~(\ref{eq:eq14})
\begin{equation}
\psi_p^{\text{dis}} = \frac{A_1}{2\rho} \mathbf{A}_c :: \mathbf{A}_c.
\label{eq:eq14}
\end{equation}
and
where $\mathbf{A}_c$ is the thermodynamic conjugate of $\mathbf{A}$, the symbol "::" means quadruple contraction, in this case the quadruple contraction between two fourth order ranked tensors produces a scalar e.g.

\[
\mathbf{A}::\mathbf{B}=\sum\limits_{i=1}^{3}\sum\limits_{j=1}^{3}\sum\limits_{k=1}^{3}\sum\limits_{l=1}^{3} A_{ijkl} B_{ijkl}.
\]
However, the yield surface equation, as shown in Eq.~(\ref{eq:yield_surface}), reduces to the von Mises yield surface without any distortion if kinematic hardening is absent (i.e., $\boldsymbol{\alpha} = \boldsymbol{0}$), even though the anisotropic part ($\mathbf{A}$) is present.

The coupled form of $\boldsymbol{\alpha}$ and $\mathbf{A}$ in Eq.~(\ref{eq:yield_surface}) causes the inconsistency between the yield surface equation and the distortional hardening rule of the complete model. This paper demonstrates how this inconsistency could be resolved. It is important to note that the total plastic free energy must be positive. Hence, distortional hardening can never happen when both isotropic hardening and kinematic hardening are absent. This is a theoretical prediction that can be made from the postulate of energy storage.
\subsection{r-model and Its Limitation \texorpdfstring{\cite{feigenbaum2008simple}}{[Feigenbaum and Dafalias, 2008]}}
A simplified model \cite{feigenbaum2008simple} developed by Feigenbaum and Dafalias introduces an orientational second-ranked tensor $\mathbf{r}$ that is used to model the distortion using a trace-type scalar term. In this model, the fourth-order tensor is not considered. The yield function of this model is simpler than the complete model \cite{feigenbaum2007directional} and is expressed in Eq.~(\ref{eq:r_model}):

\begin{equation}
f = \frac{3}{2} [1 - (\mathbf{n}^r:\mathbf{r})](\mathbf{s} - \boldsymbol{\alpha}):(\mathbf{s} - \boldsymbol{\alpha}) - k^2.
\label{eq:r_model}
\end{equation}
 The term $[1 - (\mathbf{n}^r:\mathbf{r})]$ introduces directionality to the yield surface, where,$\mathbf{n^r}$ is the normalized ($\boldsymbol{s-\alpha}$) in the radial direction as defined in \cref{eq:nprime}  and  $\mathbf{r}$ is a second order tensor-valued internal variable describing the preferred direction of distortion. As a result, the kinematic hardening has been decoupled from distortional hardening in this model.

\noindent The assumption regarding energy storage and release remains the same as the complete model discussed in \autoref{sec:intro}, where the anisotropic part of the free energy is the algebraic sum of kinematic part of free energy and distortional part of the free energy. But if that simplification is made, mentioned in Eq.~(\ref{eq:psi_reduced}), it makes it necessary to decouple kinematic hardening from distortion in the yield criterion. Feigenbaum and Dafalias~\cite{feigenbaum2008simple} realized this implication and suggested that the underlying physics requires an uncoupled consideration instead of the coupled form, which motivated them to propose this ‘$r$-model’ where these two phenomena have been decoupled.

\noindent Because of this simplification, it is possible to model distortion and translation of the yield surface separately and simultaneously, which was a major issue in the complete model. The mathematical form of the yield surface in Eq.~(\ref{eq:r_model}) is motivated by the experiments of Ishikawa~\cite{ishikawa1997subsequent}, which suggest that the yield surface starts as a ‘von Mises’ hyper-circle, and the scalar term $(\mathbf{n^r} : \mathbf{r})$, obtained from double contraction of two second rank tensor, sharpens the yield surface in the direction of loading. However, the flattening in the opposite side of the yield surface, as observed in the experiments~\cite{wu1991experimental,phillips1974some,mccomb1960some,naghdi1958experimental}, cannot be modeled using this ‘\textit{r}-model'. A yield surface model with a fourth-order anisotropic tensor could provide an effective means to capture this flattening phenomenon.

\subsection{Discussion on the Limitations of Feigenbaum’s Distortional Hardening Models}

The framework of the ‘complete model’~\cite{feigenbaum2007directional} is inconsistent, which subsequently also causes inconsistency in the mathematical formulations. The inconsistency emerges due to the coupled form of the kinematic hardening and the fourth-order anisotropic internal variable in the distortion term. Besides, in the complete model, the associative plastic flow rule is used, where the yield function acts as a plastic potential. The rationale behind using a decoupled formulation of internal variables for the plastic free energy potential, yet employing a coupled formulation for the internal variables within the plastic potential, is not explicitly evident. Because of the coupled form of the distortional term with the kinematic hardening variable, in the absence of kinematic hardening, distortion of the yield surface cannot occur, although the distortional part of the plastic free energy and the anisotropy can exist.

\noindent This inconsistency is resolved in the ‘\textit{r}-model’~\cite{feigenbaum2008simple}, as it allows distortion of the yield surface even in the absence of kinematic hardening. However, this ‘\textit{r}-model’~\cite{feigenbaum2008simple} does not have the capability to address flattening on the opposite side of the yield surface because of distortional hardening, and this flattening feature can only be captured if a fourth-order anisotropic term similar to \cref{eq:A_dot} is present in the mathematical framework of the model. 

\noindent To resolve these issues, a modified distortional hardening model is proposed that is devoid of the kinematic hardening internal variable in the distortional term but retains a fourth-order anisotropic term similar to $\mathbf{A}$. This modification allows the distortion of the yield surface even in the absence of kinematic hardening, i.e., $\boldsymbol{\alpha}=\mathbf{0}$, as well as captures the flattening on the opposite side of the yield surface. Hence, both the inconsistency of the complete model as well as the limitation of the ‘\textit{r}-model' can be overcome in the modified distortional hardening model.

\section{Development of a Modified Distortional Hardening Model}

The proposed model follows the same mathematical structure as that of the ‘complete model' except for the form of the yield surface. To maintain coherence with the proposed assumption in the ‘complete model' \cite{feigenbaum2007directional}, regarding energy storage and release, a modified version of the yield surface is presented here. The mathematical formulation of the model is developed following the same framework as discussed in \autoref{sec:2}. The energy storage assumption is the same as it is in the ‘complete model' \cite{feigenbaum2007directional}, but the form of the yield surface given in Eq.~(\ref{eq:yield_surface}) and Eq.~(\ref{eq:r_model}) is modified.

\subsection{Modification in the Expression of Anisotropy and the Proposed Yield Surface}

The ‘complete model' considers three internal variables ($k$, $\boldsymbol{\alpha}$, $\mathbf{A}$) as shown in Eq.~(\ref{eq:yield_surface}). Here, $k$ is the isotropic hardening, $\boldsymbol{\alpha}$ is the rank two deviatoric back-stress tensor, and $\mathbf{A}$ is the anisotropic fourth-order tensor-valued internal variable that models the distortion of the yield surface. 

\noindent To address this issue of the model as discussed in \autoref{sec:intro}, the distortional term $\mathbf{H} = \mathbf{H}_0 + (\mathbf{n}^r:\boldsymbol{\alpha})\mathbf{A}$ of the Feigenbaum and Dafalias model needs to be decoupled from the back-stress ($\boldsymbol{\alpha}$) so that distortional hardening can take place combined with isotropic hardening in the absence of kinematic hardening. 

\noindent Hence, an orientational tensor term ($\mathbf{r}$) is borrowed from Feigenbaum and Dafalias's ‘\textit{r}-model'\cite{feigenbaum2008simple} to introduce it in the distortional term $\mathbf{H}$ by replacing $\boldsymbol{\alpha}$ with $\mathbf{r}$. The modified distortional term is now $\mathbf{H} = \mathbf{H}_0 + (\mathbf{n}^r:\mathbf{r})\mathbf{A}$, and the modified form of the yield surface equation follows Eq.~(\ref{eq:modified_yield}) by introducing this distortional term:

\begin{equation}
f = (\mathbf{s} - \boldsymbol{\alpha}) : \{\mathbf{H}_0 + (\mathbf{n^r} : \mathbf{r}) \mathbf{A}\} : (\mathbf{s} - \boldsymbol{\alpha}) - k^2 = 0.
\label{eq:modified_yield}
\end{equation}
However, as this modified yield surface equation consists of four internal variables ($k$, $\boldsymbol{\alpha}$, $\mathbf{A}$, $\mathbf{r}$), it is not possible to obtain the evolution equation for all four internal variables (see Eqs.~(\ref{eq:group_k}--\ref{eq:k_rate}) in \autoref{sec:3.2}). The dissipation inequality can only give evolution equations for a maximum of three internal variables, as a result, the fourth one becomes arbitrary. 

\noindent Hence, the fourth-order tensor $\mathbf{A}$ is defined such that $\mathbf{A} = \mathbf{r} \otimes \mathbf{r}$ in the modified model. With this definition of anisotropic tensor, the new yield surface having now only three internal variables assumes the following form given in Eq.~(\ref{eq:new_yield}):

\begin{equation}
f = (\mathbf{s} - \boldsymbol{\alpha}):\{\mathbf{H}_0 + (\mathbf{n}^r:\mathbf{r})(\mathbf{r} \otimes \mathbf{r})\}:(\mathbf{s} - \boldsymbol{\alpha}) - k^2 = 0.
\label{eq:new_yield}
\end{equation}
To clearly highlight the modification introduced in the proposed model, the yield functions of the ‘complete model' and the ‘\( r \)-model', aligned for comparison:

\begin{align*}
\textbf{Complete model:} \quad & f = (\mathbf{s} - \boldsymbol{\alpha}):\{\mathbf{H}_0 + (\mathbf{n^r}:\boldsymbol{\alpha})\mathbf{A}\}:(\mathbf{s} - \boldsymbol{\alpha}) - k^2 = 0. \\
\textbf{\( \boldsymbol{r} \)-model:} \quad & f = \frac{3}{2} [1 - (\mathbf{n}^r:\mathbf{r})](\mathbf{s} - \boldsymbol{\alpha}):(\mathbf{s} - \boldsymbol{\alpha}) - k^2=0. \\
\textbf{Proposed model:} \quad & f = (\boldsymbol{s} - \boldsymbol{\alpha}) : \left\{ \mathbf{H}_0 + (\boldsymbol{n}^r : \boldsymbol{r})(\boldsymbol{r} \otimes \boldsymbol{r}) \right\} : (\boldsymbol{s} - \boldsymbol{\alpha}) - k^2 = 0.
\end{align*}
As shown above, the proposed model simplifies the fourth-order anisotropic tensor \( \mathbf{A} \) by directly linking it to the direction vector \( \boldsymbol{r} \), thereby reducing the number of internal variables and enhancing clarity in the formulation.

\noindent The hardening rules, flow rule, and the plastic modulus of this proposed model are expressed in \autoref{sec:3.2}.
\subsection{Hardening Rules based on the Proposed Model}
\label{sec:3.2}

Based on the proposed form of the yield surface in Eq.~(\ref{eq:new_yield}), the evolution equations for the model, along with the relevant expressions, are derived below in accordance with the second law of thermodynamics.

\subsubsection*{3.2.1 Expression of $\frac{\partial f}{\partial \boldsymbol{\sigma}}$ and Flow Rule}
\label{sec:3.2.1}

The yield surface is assumed to be convex and closed, which can be mathematically represented by the expression given in Eq.~(\ref{eq:new_yield}). The following results are helpful for the calculations:

\[
\mathbf{H}_0 : (\mathbf{s} - \boldsymbol{\alpha}) = \frac{3}{2} (\mathbf{s} - \boldsymbol{\alpha}); \quad 
\mathbf{H}_0 : \mathbf{n}^r = \frac{3}{2} \mathbf{n}^r; \quad 
\mathbf{n}^r : \mathbf{H}_0 : \mathbf{n}^r = \frac{3}{2}.
\]
Associative flow rule is assumed, and the flow rule is represented by the expression given in Eq.~(\ref{eq:flow_rule}). Taking the derivative of the yield function in Eq.~(\ref{eq:new_yield}) with respect to the stress tensor, the gradient of the yield function is obtained as shown in Eq.~(\ref{eq:df_dsigma}):

\begin{equation}
\frac{\partial f}{\partial \boldsymbol{\sigma}} = ||\mathbf{s} - \boldsymbol{\alpha}||\{3 \mathbf{n}^r + 2 (\mathbf{n}^r:\mathbf{r}) (\mathbf{r} \otimes \mathbf{r}) : \mathbf{n}^r
+ [\mathbf{n}^r : (\mathbf{r} \otimes \mathbf{r})] : \mathbf{n}^r(\mathbf{r} - (\mathbf{n}^r:\mathbf{r})\mathbf{n}^r)\}.
\label{eq:df_dsigma}
\end{equation}

\subsubsection*{3.2.2 Derivation of Hardening Rules}
\label{sec:3.2.2}

The evolution equation of the internal variables is formulated to ensure consistency with the second law of thermodynamics. The existence of the free energy per unit mass $\psi$ is assumed where $\psi$ can be additively decomposed into elastic part and plastic part. The second law of thermodynamics gives rise to the Clausius-Duhem dissipation inequality given in \cref{eq:clausius_duhem}:

\begin{equation}
\boldsymbol{\sigma} : \dot{\boldsymbol{\varepsilon}}_p - \rho \dot{\mathbf{\psi}}_p \geq 0.
\label{eq:clausius_duhem}
\end{equation}
Here, $\dot{\boldsymbol{\varepsilon}}_p$ represents the rate of plastic strain and $\rho$ is the mass density. Substituting Eq.~(\ref{eq:flow_rule}) and Eq.~(\ref{eq:df_dsigma}) into Eq.~(\ref{eq:clausius_duhem}), results into Eq.~(\ref{eq:evolution_equation}):

\begin{equation}
\lambda\, [k^2 + ||\mathbf{s} - \boldsymbol{\alpha}||^2\bigg(\frac{3}{2} + (\mathbf{n}^r:\mathbf{r})\mathbf{n}^r: (\mathbf{r} \otimes \mathbf{r}) : \mathbf{n}^r\bigg)] + \boldsymbol{\alpha} : \dot{\boldsymbol{\varepsilon}}_p - \rho \dot{\psi}_p \geq 0.
\label{eq:evolution_equation}
\end{equation}
To find the evolution equation, the exact expression for the plastic part of the free energy needs to be assumed. It is assumed that the plastic part of the free energy has two parts, an isotropic part and an anisotropic part. The anisotropic part of the free energy can further be decomposed into kinematic and distortional parts. Further assumptions about energy storage are used, which states that the postulate of Feigenbaum and Dafalias \cite{feigenbaum2007directional} on energy storage is used, which states that, isotropic and kinematic part of the free
energy correspond to energy increase by increasing the dislocation density whereas the
distortional part of the free energy corresponds to releasing energy by reorienting the
material grain to release residual stress and attain a lower state of energy. The plastic part of the free energy $\psi_p$ assumes the same form as in 
Eq.~(\ref{eq:psi_p}), and the anisotropic component of the plastic part of the free energy $\psi_p^{\text{ani}}$ assumes the same form as in Eq.~(\ref{eq:psi_ani}). The specific form of free energy is assumed to be positive definite. Following the work of Chaboche \cite{lemaitre1993mechanics} and Dafalias \cite{dafalias2003simple}, the following form of the free energy functions are assumed and expressed in Eqs.~(\ref{eq:psi_iso})--(\ref{eq:psi_dis}). From the principle of thermodynamics, it is assumed that there exists a thermodynamic conjugate variable for every internal variable. The superposition of internal varriables for different parameters is done, following the same spirit and rational by Chaboche \cite{lemaitre1993mechanics}. In accordance with the kinematic hardening framework proposed by Chaboche~\cite{lemaitre1993mechanics}, the evolution of backstress is expressed as a superposition of internal variables, each governed by independent material parameters. This superposition improves the representation of nonlinear kinematic hardening over a broader strain range, as described in ~\cite{lemaitre1993mechanics} (pp.~233--234). There, the total backstress is given by \( \mathbf{X} = \sum_{\ell} \mathbf{X}_\ell \), with each component following its own evolution law of the form \( \mathrm{d} \mathbf{X}_\ell = \tfrac{2}{3} C_\ell \, \mathrm{d} \boldsymbol{\varepsilon}^p - \gamma_\ell \, \mathbf{X}_\ell \, \mathrm{d}p \).

In this work, a similar decomposition is applied to the plastic part of the free energy, expressed in Eqs.~(21)--(23), where the internal variables \( k_c \), \( \boldsymbol{\alpha}_c \), and \( \mathbf{r}_c \) correspond respectively to isotropic, kinematic, and distortional contributions. Each term is associated with its own set of material parameters (\( \kappa_i \), \( a_i \), and \( A_i \)). The thermodynamic conjugates for $k$, $\boldsymbol{\alpha}$, and $\mathbf{r}$ are $k_c$, $\boldsymbol{\alpha}_c$, and $\mathbf{r}_c$. 

\begin{equation}
\psi_p^{\text{iso}} = \sum_{i=1}^{l} \frac{\kappa_i}{2\rho} k_{c}^2,
\label{eq:psi_iso}
\end{equation}

\begin{equation}
\psi_p^{\text{kin}} = \sum_{i=1}^{l} \frac{a_i}{2\rho} \boldsymbol{{\alpha_{c}}} : \boldsymbol{{\alpha_{c}}},
\label{eq:psi_kin}
\end{equation}

\begin{equation}
\psi_p^{\text{dis}} = \sum_{i=1}^{l} \frac{A_i}{2\rho} \mathbf{r_{c}} : \mathbf{r_{c}}.
\label{eq:psi_dis}
\end{equation}
To keep the calculation simple and the number of parameters to the minimum, only the first term is considered in Eqs.~(\ref{eq:psi_iso})--(\ref{eq:psi_dis}), i.e., $l=1$, and no superposition is considered. The final expressions are given in Eqs.~(\ref{eq:psi_iso_final})--(\ref{eq:psi_dis_final}):

\begin{equation}
\psi_p^{\text{iso}} = \frac{\kappa_1}{2\rho} k_c^2,
\label{eq:psi_iso_final}
\end{equation}

\begin{equation}
\psi_p^{\text{kin}} = \frac{a_1}{2\rho} \boldsymbol{\alpha}_c : \boldsymbol{\alpha}_c,
\label{eq:psi_kin_final}
\end{equation}

\begin{equation}
\psi_p^{\text{dis}} = \frac{A_1}{2\rho} \mathbf{r}_c : \mathbf{r}_c.
\label{eq:psi_dis_final}
\end{equation}
In Eqs.~(\ref{eq:psi_iso_final})--(\ref{eq:psi_dis_final}), the material constants $\kappa_1$, $a_1$, and $A_1$ are non-negative material constants. They are related to their conjugates by the following relationships, as shown in Eqs.~(\ref{eq:conjugate_k})--(\ref{eq:conjugate_r}):

\begin{equation}
k = \rho \frac{\partial \psi_p^{\text{iso}}}{\partial k_c} = \kappa_1 k_c,
\label{eq:conjugate_k}
\end{equation}

\begin{equation}
\boldsymbol{\alpha} = \rho \frac{\partial \psi_p^{\text{kin}}}{\partial \boldsymbol{\alpha}_c} = a_1 \boldsymbol{\alpha}_c,
\label{eq:conjugate_alpha}
\end{equation}

\begin{equation}
\mathbf{r} = \rho \frac{\partial \psi_p^{\text{dis}}}{\partial \mathbf{r}_c} = A_1 \mathbf{r}_c.
\label{eq:conjugate_r}
\end{equation}
Substituting the expression of Eq.~(\ref{eq:psi_total}) in Eq.~(\ref{eq:evolution_equation}) and then using the expressions given in Eqs.~(\ref{eq:psi_iso_final})--(\ref{eq:conjugate_r}) leads to the Clausius-Duhem inequality, presented in Eq.~(\ref{eq:inequality}) that is required to be satisfied at every stage of loading:

\begin{equation}
\lambda \big[k^2 + ||\mathbf{s} - \boldsymbol{\alpha}||^2 \bigg(\frac{3}{2}+(\mathbf{n}^r:\mathbf{r})\mathbf{n}^r:(\mathbf{r} \otimes \mathbf{r}):\mathbf{n}^r\bigg)\big] 
+ \boldsymbol{\alpha} : \bigg(\dot{\boldsymbol{\varepsilon}}_p - \frac{1}{a_1} \dot{\boldsymbol{\alpha}} \bigg) 
- \frac{1}{\kappa_1} k \dot{k} + \frac{1}{A_1} \mathbf{r} : \dot{\mathbf{r}} \geq 0.
\label{eq:inequality}
\end{equation}
Grouping the terms in Eq.~(\ref{eq:inequality}), which contain the same internal variables, and imposing the condition that each group must be non-negative individually, as shown in Eqs.~(\ref{eq:group_k})--(\ref{eq:group_r}), ensures the satisfaction of inequality presented in Eq.~(\ref{eq:inequality}).

\begin{equation}
k \bigg(\lambda k - \frac{1}{\kappa_1} \dot{k} \bigg) \geq 0,
\label{eq:group_k}
\end{equation}

\begin{equation}
\boldsymbol{\alpha} : \bigg(\dot{\boldsymbol{\varepsilon}}_p - \frac{1}{a_1} \dot{\boldsymbol{\alpha}} \bigg) \geq 0,
\label{eq:group_alpha}
\end{equation}

\begin{equation}
\lambda  ||\mathbf{s} - \boldsymbol{\alpha}||^2 \bigg(\frac{3}{2} + (\mathbf{n}^r:\mathbf{r})\mathbf{n}^r:(\mathbf{r} \otimes \mathbf{r}):\mathbf{n}^r\bigg) + \frac{1}{A_1} \mathbf{r} : \dot{\mathbf{r}} \geq 0.
\label{eq:group_r}
\end{equation}
To satisfy the inequalities in Eqs.~(\ref{eq:group_k})--(\ref{eq:group_r}), the following form of the functions are assumed, and the rate equation for each internal variable is derived, as shown in Eqs.~(\ref{eq:k_rate})--(\ref{eq:r_dot}):

\begin{equation}
\lambda k - \frac{1}{\kappa_1} \dot{k}= \lambda \kappa_2k^2\implies \dot{k} = \lambda k \kappa_{1} (1 - \kappa_2 k),
\label{eq:k_rate}
\end{equation}

\begin{equation}
\dot{\boldsymbol{\varepsilon}_p} - \frac{1}{a_1} \dot{\boldsymbol{\alpha}} = \lambda \left\| \frac{\partial f}{\partial \boldsymbol{\sigma}} \right\| a_2 \boldsymbol{\alpha} 
\quad \implies \quad 
\dot{\boldsymbol{\alpha}} = \lambda \left\| \frac{\partial f}{\partial \boldsymbol{\sigma}} \right\| a_1 \left( \mathbf{n} - a_2 \boldsymbol{\alpha} \right),
\label{eq:alpha_rate}
\end{equation}
\hspace{50 pt}$\lambda\, ||\mathbf{s}-\boldsymbol{\alpha}||^2 \Bigl[(\mathbf{n^r}: \mathbf{r})\, \mathbf{n^r} : (\mathbf{r} \otimes \mathbf{r}) : \mathbf{n^r}\Bigr] + \frac{1}{A_1}\, \mathbf{r} : \boldsymbol{\dot{r}} 
= -\frac{3}{2}\lambda\, ||\mathbf{s}-\boldsymbol{\alpha}||^2\, A_2\, (\mathbf{r} : \mathbf{r}),\\ $ 
\begin{equation}
\Longrightarrow \quad \boldsymbol{\dot{r}} = -\lambda\, A_1\, ||\mathbf{s}-\boldsymbol{\alpha}||^2 \Bigl[(\mathbf{n^r} :\mathbf{r})^2\, \mathbf{n^r} + \frac{3}{2}\, A_2\, \mathbf{r}\Bigr].
\label{eq:r_dot}
\end{equation}
where \(\kappa_2\), \(\alpha_2\), and \(A_2\) are material parameters.
The evolution equation for the anisotropic term $\mathbf{A}$ can be written as shown in \cref{eq:eqA_rate} 
\begin{equation}
\dot{\mathbf{A}} = \dot{\mathbf{r}} \otimes \mathbf{r} + \mathbf{r} \otimes \dot{\mathbf{r}}.
\label{eq:eqA_rate}
\end{equation}
Substituting the expression obtained in \cref{eq:r_dot} into \cref{eq:eqA_rate}
\begin{equation}
\dot{\mathbf{A}} = -\lambda\, A_1\, ||\mathbf{s} - \boldsymbol{\alpha}||^2 
\left[ (\mathbf{n}^r : \mathbf{r})\, \mathbf{n}^r \otimes \mathbf{n}^r 
+ \frac{3}{2} A_2\, (\mathbf{r} \otimes \mathbf{r}) \right],
\label{eq:A_rate3}
\end{equation}

\begin{equation}
\dot{\mathbf{A}} = -\lambda\, A_1\, ||\mathbf{s} - \boldsymbol{\alpha}||^2 
\left[ (\mathbf{n}^r : \mathbf{r})\, \mathbf{n}^r \otimes \mathbf{n}^r 
+ \frac{3}{2} A_2\, \mathbf{A}  \right].
\label{eq:A_rate3}
\end{equation}

\cref{eq:A_rate3} has the same mathematical structure as \cref{eq:A_dot}. Based on the above mathematical analysis, it can be concluded that in the proposed model with the term $\mathbf{A}$= $\mathbf{r} \otimes \mathbf{r}$ possesses all the necessary characteristics to simultaneously capture both the flattening and the sharpening of the yield surface like the fourth-order anisotropic term $\mathbf{A}$ does in the ‘complete model’ \cite{feigenbaum2007directional}.

\noindent The rate equations obtained are of the evanescent memory type, i.e., these are the Armstrong-Frederick model type equations. This means the internal variables have an upper limit on their evolution and are path-dependent. For Eq.~(\ref{eq:r_dot}) to satisfy Eq.~(\ref{eq:group_r}), the necessary condition is given in Eq.~(\ref{eq:constraint_r}):

\begin{equation}
A_2 (\mathbf{r} : \mathbf{r}) \leq 1 .\quad \forall \, \mathbf{r}
\label{eq:constraint_r}
\end{equation}
From Eqs.~(\ref{eq:alpha_rate}) and (\ref{eq:r_dot}), it can be seen that $\boldsymbol{\alpha}$ evolves in the direction of $\mathbf{n}$ and $\mathbf{r}$ evolves in the direction of $\mathbf{n}^r$. All three of the evolution equations involve two material parameters, which means a total of six material parameters is required. Evolution of $\mathbf{r}$ is such that it must always satisfy Eq.~(\ref{eq:constraint_r}).

\subsubsection*{3.2.3 Plastic Modulus}

The final step is to find the loading index $\lambda$. The loading index is required to find the plastic strain rate using the flow rule given in Eq.~(\ref{eq:flow_rule}). During plastic loading, the consistency condition $\dot{f} = 0$ must be obeyed, which results in the following expression in Eqs.~(\ref{eq:consistency1}--\ref{eq:consistency2}):

\begin{equation}
\frac{\partial f}{\partial \boldsymbol{\sigma}} : \dot{\boldsymbol{\sigma}} + \frac{\partial f}{\partial \boldsymbol{\alpha}} : \dot{\boldsymbol{\alpha}} + \frac{\partial f}{\partial k} \dot{k} + \frac{\partial f}{\partial \mathbf{r}} : \dot{\mathbf{r}} = 0,
\label{eq:consistency1}
\end{equation}

\begin{equation}
\frac{\partial f}{\partial \boldsymbol{\sigma}} : \dot{\boldsymbol{\sigma}} 
+ \lambda \left( 
\frac{\partial f}{\partial \boldsymbol{\alpha}} : \bar{\boldsymbol{\alpha}} 
+ \frac{\partial f}{\partial k} \bar{k} 
+ \frac{\partial f}{\partial \mathbf{r}} : \bar{\mathbf{r}} 
\right) = 0.
\label{eq:consistency2}
\end{equation}
Taking the derivative of the yield function given in Eq.~(\ref{eq:new_yield}) with respect to the back-stress results in the following expression, given in Eq.~(\ref{eq:df_dalpha}):

\begin{equation}
\frac{\partial f(\boldsymbol{\sigma}, \boldsymbol{\alpha}, \mathbf{r}, k)}{\partial \boldsymbol{\alpha}} = -\frac{\partial f(\boldsymbol{\sigma}, \boldsymbol{\alpha}, \mathbf{r}, k)}{\partial \boldsymbol{\sigma}}.
\label{eq:df_dalpha}
\end{equation}

Taking the derivative of the yield function with respect to the internal variable $\mathbf{r}$ results in the following expression in Eq.~(\ref{eq:df_dr}):

\begin{equation}
\frac{\partial f}{\partial \mathbf{r}} = 3 ||\mathbf{s} - \boldsymbol{\alpha}||^2 (\mathbf{n}^r:\mathbf{r})^2\, \mathbf{n}^r.
\label{eq:df_dr}
\end{equation}
Using the definition of plastic modulus given in Eq.~(\ref{eq:lambda}) and substituting Eq.~(\ref{eq:df_dalpha}) and Eq.~(\ref{eq:df_dr}) into Eq.~(\ref{eq:consistency2}) results into the expression of the plastic modulus $K_p$, as shown in Eqs.~(\ref{eq:Kp1}--\ref{eq:Kp2}):

\begin{equation}
K_p = -\frac{\partial f}{\partial \boldsymbol{\alpha}} : \bar{\boldsymbol{\alpha}} - \frac{\partial f}{\partial k} \bar{k} - \frac{\partial f}{\partial \mathbf{r}} : \bar{\mathbf{r}},
\label{eq:Kp1}
\end{equation}

\begin{equation}
K_p = 2 \kappa_1 k^2 (1 - \kappa_2 k) + \left\|\frac{\partial f}{\partial \boldsymbol{\sigma}}\right\|  a_1 \frac{\partial f}{\partial \boldsymbol{\sigma}} (\mathbf{n} - a_2 \boldsymbol{\alpha}) 
+ 3 A_1 ||\mathbf{s} - \boldsymbol{\alpha}||^4 (\mathbf{n}^r:\mathbf{r})^3 \big(\mathbf{n}^r:\mathbf{r} + \frac{3}{2} A_2\big).
\label{eq:Kp2}
\end{equation}
This completes the mathematical formulation for the proposed model.
\subsubsection*{3.2.4 Constraints on Material Parameters}

Constraints on material parameters can be obtained by exploiting the positiveness and convexity requirements at the limiting value of the internal variables. Setting the rates in Eq.~(\ref{eq:k_rate}) to Eq.~(\ref{eq:r_dot}) to zero, the respective results are expressed in Eqs.~(\ref{eq:limiting_k}--\ref{eq:limiting_r}):

\begin{equation}
k^l = \frac{1}{\kappa_2},
\label{eq:limiting_k}
\end{equation}

\begin{equation}
\boldsymbol{\alpha}^l = \frac{\mathbf{n}^l}{a_2},
\label{eq:limiting_alpha}
\end{equation}

\begin{equation}
\mathbf{r}^l = -\frac{3}{2} A_2 \mathbf{n^r}^l.
\label{eq:limiting_r}
\end{equation}
Here, $k^l$, $\boldsymbol{\alpha}^l$, and $\mathbf{r}^l$ are the limiting values of the hardening variables $k$, $\boldsymbol{\alpha}$, and $\mathbf{r}$. Substituting Eq.~(\ref{eq:limiting_r}) into Eq.~(\ref{eq:constraint_r}) gives the constraint on the material parameter $A_2$, as shown in Eq.~(\ref{eq:constraint_A2_1}):

\begin{equation}
A_2 \leq {\sqrt[3]{4/9}}.
\label{eq:constraint_A2_1}
\end{equation}
Eq.~(\ref{eq:constraint_A2_1}): puts a constraint on the upper bound of $A_2$. Another way to find and cross-check the limit is by substituting Eqs.~(\ref{eq:limiting_k}--\ref{eq:limiting_r}) into the expression of the gradient of the yield surface at the limiting condition. Substituting the limiting values of the internal variables obtained in Eqs.~(\ref{eq:limiting_k}--\ref{eq:limiting_r}) into the expression of the gradient of the yield surface given in Eq.~(\ref{eq:df_dsigma}), the limiting value of the gradient is obtained and given in Eq.~(\ref{eq:gradient_limiting}):

\begin{equation}
\frac{\partial f}{\partial \boldsymbol{\sigma}^l} = \left\|\mathbf{s}^l - \boldsymbol{\alpha}^l \right\|\bigg[3 - \frac{27}{4} A_2^3 \bigg] \mathbf{n^r}^l
\label{eq:gradient_limiting}.
\end{equation}
Taking the Frobenius norm of Eq.~(\ref{eq:gradient_limiting}) gives the magnitude of the gradient at the limiting condition, given in Eq.~(\ref{eq:gradient_magnitude}):

\begin{equation}
\left\|\frac{\partial f}{\partial \boldsymbol{\sigma}^l} \right\| = \left\|\mathbf{s}^l - \boldsymbol{\alpha}^l\right\| \bigg[3 - \frac{27}{4} A_2^3 \bigg] > 0.
\label{eq:gradient_magnitude}
\end{equation}
Eq.~(\ref{eq:gradient_magnitude}) puts the following constraint on the material parameter $A_2$, as shown in Eq.~(\ref{eq:constraint_A2_2}):

\begin{equation}
A_2 \leq {\sqrt[3]{4/9}}.
\label{eq:constraint_A2_2}
\end{equation}
Remarkably, the thermodynamic approach and the convexity requirement of the yield surface both agree upon the restriction put on the material constant. In addition, the lower bound for $A_2$ can be calculated from the requirement of positive definiteness of the equivalent stress. This condition, evaluated at the limiting case of $\mathbf{n}^r$ and $\boldsymbol{\alpha}$, leads to the expression given in Eq.~(\ref{eq:positive_definiteness}).

\begin{equation}
(\mathbf{s} - \boldsymbol{\alpha}) : \mathbf{H} : (\mathbf{s} - \boldsymbol{\alpha}) = \frac{3}{2} ||\mathbf{s}^l - \boldsymbol{\alpha}^l||^2 + (\mathbf{n^r}^l : \mathbf{r}^l)^3\, ||\mathbf{s}^l - \boldsymbol{\alpha}^l||^2 \geq 0.
\label{eq:positive_definiteness}
\end{equation}
Upon substitution of Eqs.~(\ref{eq:limiting_k}), (\ref{eq:limiting_alpha}), and (\ref{eq:limiting_r}) into Eq.~(\ref{eq:positive_definiteness}), the constraint on the material parameter $A_2$ is obtained, and the following relations are given in Eqs.~(\ref{eq:A2_constraint1}) and (\ref{eq:A2_constraint2}):

\begin{equation}
\frac{3}{2} ||\mathbf{s^l} - \boldsymbol{\alpha^l}||^2 - ||\mathbf{s^l} - \boldsymbol{\alpha^l}||^2 \,\frac{27}{8} A_2^3\, (\mathbf{n^r} : \mathbf{n^r}^l)^3 \geq 0,
\label{eq:A2_constraint1}
\end{equation}

\begin{equation}
\frac{3}{2} ||\mathbf{s^l} - \boldsymbol{\alpha^l}||^2 \bigg[1 - \frac{9}{4} A_2^3\, (\mathbf{n^r} : \mathbf{n^r}^l)^3 \bigg] \geq 0.
\label{eq:A2_constraint2}
\end{equation}
Here, lets assume two possibilities $\mathbf{n^r} : \mathbf{n^r}^l = \pm 1$. That means they can act in the same direction or in the opposite direction. If they work in the opposite direction ($\mathbf{n^r} : \mathbf{n^r}^l = -1$), the following is the constraint on material parameter $A_2$, as shown in Eq.~(\ref{eq:A2_same_direction}):

\begin{equation}
A_2 \geq{\sqrt[3]{-4/9}}.
\label{eq:A2_same_direction}
\end{equation}
Which puts a lower bound on $A_2$. However, if $\mathbf{n}^r$ and $\mathbf{n}^{rl}$ are in the same direction (i.e., $\mathbf{n}^r : \mathbf{n}^{r^l} = +1$), then Eqs.~(\ref{eq:constraint_A2_1}) and (\ref{eq:constraint_A2_2}) are recovered.

\noindent Another criterion that needs to be satisfied is the positive definiteness of $\psi_p^{\text{ani}}$ at the limiting value. Because from Eqs.~(\ref{eq:psi_p}) and (\ref{eq:psi_ani}), it becomes apparent that it is a sufficient condition to ensure the positive definiteness of $\psi_p$. Substituting Eq.~(\ref{eq:psi_kin_final}) and Eq.~(\ref{eq:psi_dis_final}) into Eq.~(\ref{eq:psi_ani}) at the limiting state results in the following expression in Eq.~(\ref{eq:psi_ani_limiting}):

\begin{equation}
\psi_p^{\text{ani}} = \frac{1}{\rho} \bigg[\frac{1}{a_2} \boldsymbol{\alpha}^l : \boldsymbol{\alpha}^l - \frac{1}{A_1} \mathbf{r}^l : \mathbf{r}^l \bigg].
\label{eq:psi_ani_limiting}
\end{equation}
Upon substituting the limiting value found in Eqs.~(\ref{eq:limiting_alpha}) and (\ref{eq:limiting_r}) into Eq.~(\ref{eq:psi_ani_limiting}), a second material constraint is obtained, as given in Eq.~(\ref{eq:material_constraint2}):

\begin{equation}
\frac{9 a_1 a_2^2 A_2^2}{4 A_1} < 1.
\label{eq:material_constraint2}
\end{equation}
\( a_2 \), \( \kappa_2 \), and \( A_2 \) are associated with \( \boldsymbol{\alpha^l} \), \( \boldsymbol{\kappa^l} \), and \( \boldsymbol{r^l} \), respectively which are the limiting values of the internal variables. And \( a_1 \), \( \kappa_1 \), and \( A_1 \) are associated with the rate at which this limit is reached.

\section{Numerical Implementation}

The goal of this section is to show that the theoretical model can be implemented numerically using the algorithm described in the \cref{Appendix}. To show this the material is loaded uniaxially with a strain increment of $5\times10^{-5}$. The material is loaded up to a total strain of one percent.  For this numerical experiment, the values of the material parameters are as follows:

\begin{table}[H]
\centering
\caption{Material Parameters used for numerical experiment}
\label{tab:parameters}
\begin{tabular}{lc}
\toprule
\textbf{Parameter} & \textbf{Value} \\
\midrule
$E$         & 196687 MPa \\
$\nu$       & 0.28 \\
$k_{initial}$          & 128 MPa \\
$\kappa_1$       & 26500\\
$\kappa_2$       & 0.003 MPa$^{-1}$,\\
$a_1$       & 828840\\
$a_2$       & 0.011 MPa$^{-1}$, \\
$A_1$    & 6.28 MPa$^{-2}$, \\
$A_2$    & 0.046\\
$dX$     & 0.00005 \\
\bottomrule
\end{tabular}
\end{table}

\noindent
The material parameter values are selected in a pedagogical manner, respecting the constraints imposed by the theoretical framework. The evolution of the internal variables is governed by the following equations:

\begin{equation}
\left.
\begin{aligned}
\dot{k} &= \lambda k \kappa_{1} (1 - \kappa_2 k), \quad
\dot{\boldsymbol{\alpha}} = \lambda \left\| \frac{\partial f}{\partial \boldsymbol{\sigma}} \right\| a_1 \left( \mathbf{n} - a_2 \boldsymbol{\alpha} \right), \\
\dot{\mathbf{r}} &= -\lambda\, A_1\, \|\mathbf{s} - \boldsymbol{\alpha}\|^2 
\left[ (\mathbf{n}^r : \mathbf{r})^2\, \mathbf{n}^r + \frac{3}{2} A_2\, \mathbf{r} \right].
\end{aligned}
\right\}
\label{eq:evolution_equations}
\end{equation}

\vspace{1em}

\noindent
The following figures in \cref{fig:evolution_variables} illustrate the evolution of the internal variables involved in the constitutive model:
\begin{figure}[H]
    \centering
    \subfigure[]{\includegraphics[width=0.33\linewidth,height=4.9cm]{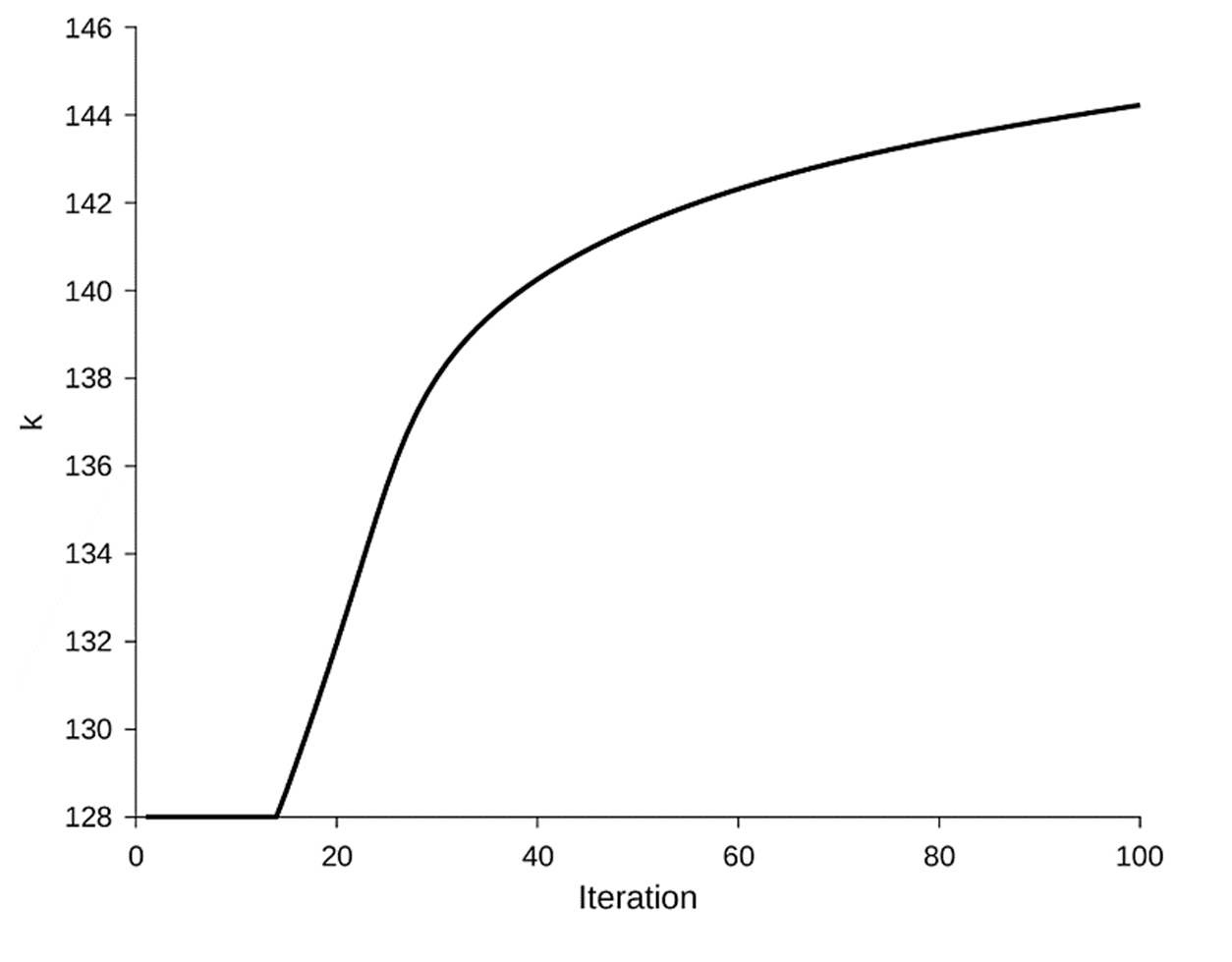}}
	\subfigure[]{\includegraphics[width=0.33\linewidth,height=5cm]{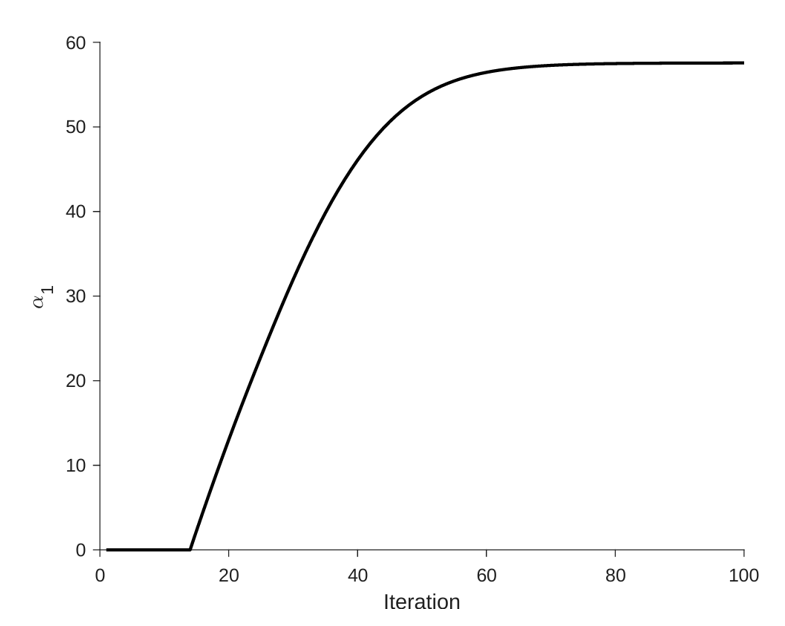}} 
    \subfigure[]{\includegraphics[width=0.33\linewidth,height=5cm]{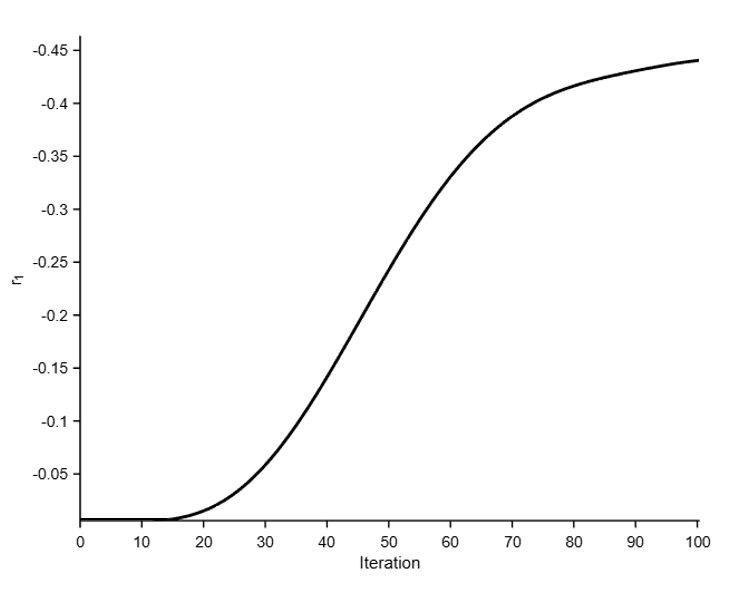}}
    \caption{Evolution of internal variables (a) $k$, (b) $\alpha_1$, and (c) $r_1$  over the loading iterations}
    \label{fig:evolution_variables}
   \end{figure}

\noindent
The evolution of the internal variables is smooth and free from numerical defects. This confirms the stability of the numerical implementation. The backstress $\alpha_1$ evolves continuously, reflecting the kinematic hardening, while $r_1$ captures the distortion effects.

\vspace{1em}

\begin{figure}[H]
    \centering
    \includegraphics[width=0.65\linewidth]{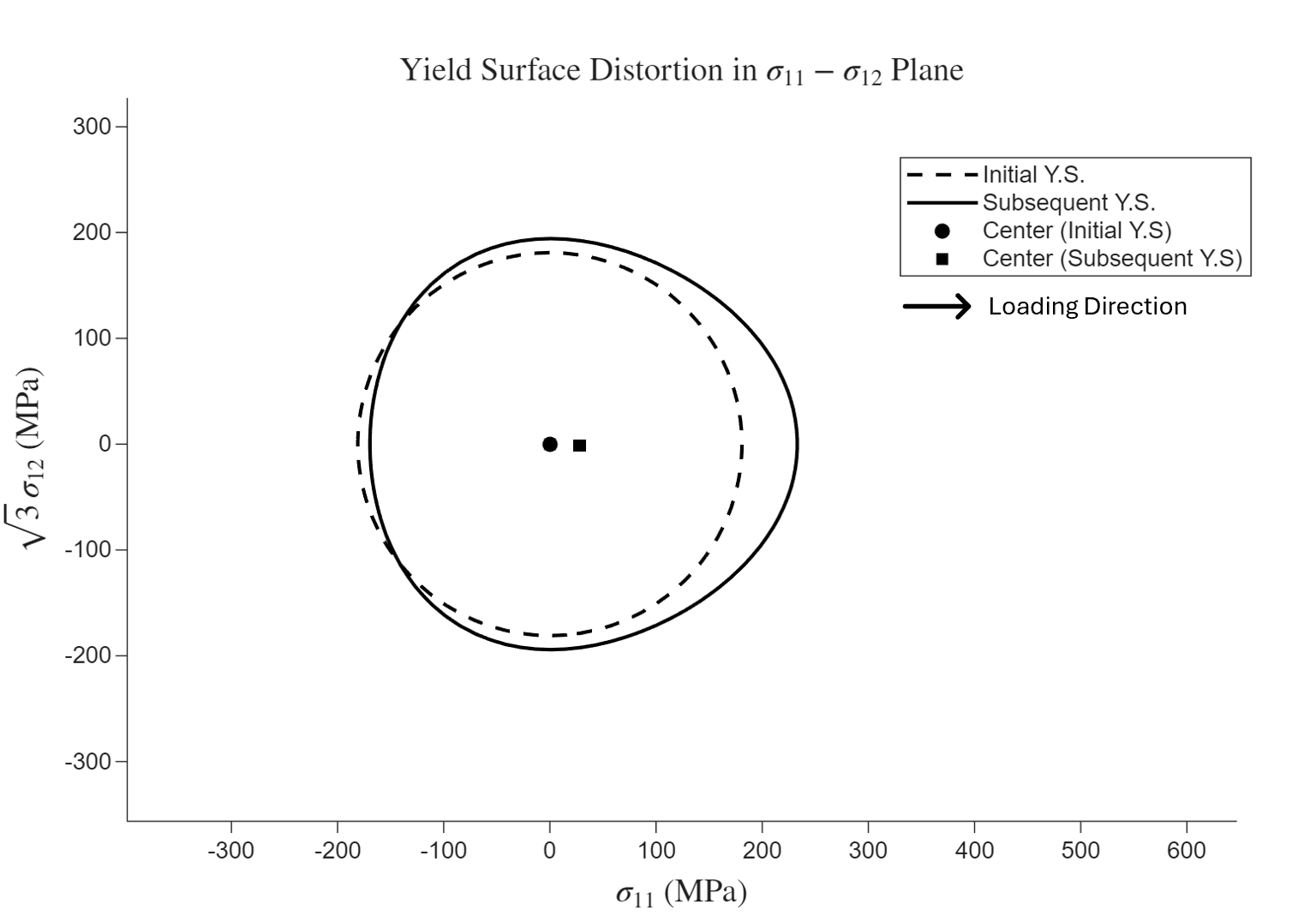}
    \caption{Evolution and distortion of the modeled Yield Surface (Y.S.) in the $\sigma_{11}$--$\sigma_{12}$ plane.}
    \label{fig:distortion}
\end{figure}

\noindent
\cref{fig:distortion} illustrates the yield surface evolution under monotonic loading. The model captures complex distortional behavior including sharpening in one direction and flattening in the opposite. Such behavior is driven by the evolution of the distortion tensor $\boldsymbol{r}$. The formulation allows independent control over translation, distortion, and rotation in the stress space, providing greater flexibility.

\section{Discussion}

In the proposed modified model, when the back-stress ($\boldsymbol{\alpha}$) is zero from Eq.~(\ref{eq:new_yield}), it can be seen that the equation of the yield surface is not of the von Mises yield surface type, unlike the ‘complete model' of Feigenbaum et al. \cite{feigenbaum2007directional}, and distortional hardening can still occur with isotropic hardening. That means, for some materials, combined isotropic hardening with distortional hardening is possible when kinematic hardening is absent. This model of the yield surface does not contradict the energy assumption made in the ‘complete model' \cite{feigenbaum2007directional} of Feigenbaum and Dafalias. Moreover, flattening of the yield surface can also be captured, unlike the ‘\textit{r}-model' of Feigenbaum and Dafalias \cite{feigenbaum2008simple}.

\noindent In Eq.~(\ref{eq:inequality}), an assumption is made pertaining to the separation of the terms that are similar and obtaining three inequalities for three internal variables from that one inequality. This approach, while not derived from general physical principles, can be interpreted as being justified under the assumption that the three dissipation inequalities correspond to independent mechanisms of energy dissipation. The assumptions made in Eqs.~(\ref{eq:k_rate}) to (\ref{eq:r_dot}) are done pedagogically so that the evolution equations follow the Armstrong-Frederick type hardening form. To increase the accuracy of the model, superposition on the hardening variables can be employed like the ‘Chaboche Hardening Rule,’ but that would introduce additional parameters into the model, and calibrating them will further add complexity.

\noindent Out of the six material parameters in the model, numerical bounds on only two material constraints, $A_1$ and $A_2$, can be calculated theoretically, as presented in this paper. The numerical implementation of the proposed model discussed in this study can be performed using the linearized integration technique proposed by Bardet and Choucair~\cite{bardet1991linearized}, which is well suited for rate-type and incremental constitutive equations under generalized loadings. This method transforms nonlinear constitutive behavior and loading constraints into a linear system of ordinary differential equations. A detailed step-by-step description of the algorithm for this model is provided in the \cref{Appendix}.

The numerical experiments presented in this study serve as a verification of the implementability of the proposed constitutive model. Using the algorithm described in the \cref{Appendix}, the model was subjected to a uniaxial loading condition with controlled strain increments. The simulation results confirm that the model is numerically stable and capable of capturing key features of elastoplastic behavior, including yield surface distortion and the evolution of internal variables. However, it must be emphasized that these results, while valuable, do not in themselves validate the predictive capability of the model. The current simulations are performed for demonstration purposes using assumed parameter values. For the model to be used as a reliable predictive tool in material modeling, it must be calibrated against experimental data. Calibration would require the identification of appropriate values for the six material parameters through fitting to, yield surface evolution data. 

\section*{Acknowledgment}
The authors gratefully acknowledge the valuable insights and constructive feedback provided by Professors H. P. Feigenbaum and Y. F. Dafalias. Their foundational work has profoundly influenced this study. 

The first and second author is especially grateful to Professor Y. F. Dafalias, whose work not only guided this study but also introduced them to the field of rational mechanics.

\section*{Funding and Competing interests}

No funds, grants, or other support was received for this study. The authors have no relevant financial or non-financial interests to disclose.
\section*{Author Contributions}
Md Mahmudur Rahman (M.M.R.), Md Mahmudul Hasan Pathik (M.M.H.P.) contributed equally to this work.

\textbf{M.M.R.} : conceptualization, formal analysis, investigation, methodology, writing—original draft, writing—review and editing; \textbf{M.M.H.P.}  : conceptualization, formal analysis, investigation, methodology, writing—original draft, writing—review and editing;\textbf{ N.I.} : investigation, supervision, writing—original draft, writing—review and editing.
All authors gave final approval for publication and agreed to be held accountable for the work performed therein.

\clearpage
\appendix
\section[Appendix: Bardet and Choucair]{Appendix: Bardet and Choucair \cite{bardet1991linearized} Numerical Procedure}
 \label{Appendix}
In this study, the linearized integration technique is formulated under the assumption of small strains, whereby rigid body rotations (spin effects) are neglected, and objective stress rates are considered equivalent to material stress increments.
The numerical procedure used by Bardet and Choucair~\cite{bardet1991linearized} can be utilized to numerically implement this work is summarized below.

\begin{enumerate}
    \item \textbf{Initialization}: Define initial values for stress $\boldsymbol{\sigma}$, strain $\boldsymbol{\varepsilon}$, internal variables $k$, $\boldsymbol{\alpha}$, and $\mathbf{r}$. If we consider the loading conditions involve either normal stresses (\(\sigma_{11}\), \(\sigma_{22}\), and \(\sigma_{33}\)), shear stresses (\(\sigma_{12}\)), or a combination of two normal stresses the stress components \(\sigma_{13}\), \(\sigma_{23}\), strain components \(\varepsilon_{13}\), \(\varepsilon_{23}\), and corresponding internal variables are zero. As a result, stress, strain, and internal variables of second rank can be simplified into (\(4 \times 1\)) vectors, and fourth-order tensors as (\(4 \times 4\)) matrices for computational efficiency.

    \item \textbf{Loading Conditions}:
   Based on the loading conditions, define $\mathbf{S}$, $\mathbf{E}$, and loading/strain increment size $dX$ depending on the choice of $\mathbf{S}$, and $\mathbf{E}$, ensuring the following equation is satisfied:
    \[
    \mathbf{S} d \boldsymbol{\sigma} + \mathbf{E}d \boldsymbol{\varepsilon} = (\mathbf{S} + \mathbf{E} \mathbf{B}) d \boldsymbol{\sigma} = \mathbf{d} \mathbf{Y}.
    \]
    where \(d\mathbf{Y} = \begin{bmatrix} 0 & 0 & 0 & dX \end{bmatrix}^T\), and \(\mathbf{B}\) is the elasto-plastic constitutive matrix (\(d \boldsymbol{\varepsilon} = \mathbf{B}d \boldsymbol{\sigma}\)). The choice of $\boldsymbol{S}$ and $\boldsymbol{E}$, determined by the loading setup, varies depending on whether the case involves stress controlled, strain controlled, or other loading scenarios \cite{bardet1991linearized,janda2017general}.

    \item \textbf{Constitutive Matrix Calculation}:
    Initially, assume elastic behavior and compute $\mathbf{B}$ using the isotropic linear elastic matrix:
    \[ \mathbf{B}^{trial}=
    \mathbf{B}^e = \frac{1}{E}\begin{bmatrix}
    1 & -\nu & -\nu & 0\\[5pt]
    -\nu & 1 & -\nu & 0\\[5pt]
    -\nu & -\nu & 1 & 0\\[5pt]
    0 & 0 & 0 & 1+\nu
    \end{bmatrix}
    \]
    where ${E}$ is Young’s modulus and $\nu$ is the Poisson’s ratio. Using $\mathbf{B}^{trial}$ as $\mathbf{B}$ in the equation at step 2, $d\boldsymbol{\sigma}$ can be calculated.

    \item \textbf{Elastic Check}:
    Compute the yield function \(f(\boldsymbol{\sigma} + {d} \boldsymbol{\sigma})\). If \(f \leq 0\), the behavior is purely elastic, then skip to step 11 to update stresses and strains. Otherwise, proceed to step 5.

    \item \textbf{Elasto-Plastic Transition}:
    \begin{itemize}
        \item Check if $f(\boldsymbol{\sigma} + d\boldsymbol{\sigma}) > 0$, then the loading is elasto-plastic. 
        \item Solve for $x$ such that $f(\boldsymbol{\sigma} + x d\boldsymbol{\sigma}) = 0$, with $0 \leq x \leq 1$ to get the purely elastic part. 
    \end{itemize}

    \item \textbf{Update Stress and Strain}:
    \begin{itemize}
    \item Update the elastic part of stress and strain using the x value:
    \[\boldsymbol{\sigma} \leftarrow \boldsymbol{\sigma} + x {d} \boldsymbol{\sigma}, \quad \boldsymbol{\varepsilon} \leftarrow \boldsymbol{\varepsilon} + x \mathbf{B}{d}\boldsymbol{\sigma} \vspace{1em}.
    \]
\item  Adjust the loading increment: \(d {Y}_4 \leftarrow (1 - x) dX\) for the remaining part which involves plastic deformation.

\end{itemize}
    \item \textbf{Calculate Plastic Modulus $K_p$}:
    Compute $K_p$ from \cref{eq:Kp2}.

    \item \textbf{Form Elasto-Plastic Constitutive Matrix $\mathbf{B}$}:
    Update \(\mathbf{B}\) using Hooke’s law and the associative flow rule:
    \[
    \mathbf{B} = \mathbf{B}^e + \frac{1}{K_p}\frac{\partial f}{\partial \sigma}\otimes \frac{\partial f}{\partial \mathbf{\sigma}}.
    \]

    \item \textbf{Solve Stress Increment}:
    Compute $d\boldsymbol{\sigma} = (\mathbf{S}+\mathbf{EB})^{-1} d\mathbf{Y}$ for the plastic deformation part.

    \item \textbf{Update Internal Variables}:
     If \(K_p \frac{\partial f}{\partial \boldsymbol{\sigma}} : \mathbf{d} \boldsymbol{\sigma} > 0\), calculate multiplier $\lambda$ using \cref{eq:lambda} and update internal variables using hardening rules derived in \cref{sec:3.2.2}.

    \item \textbf{Update Stress and Strain}:
    Update stress $\boldsymbol{\sigma} = \boldsymbol{\sigma} + d\boldsymbol{\sigma}$ and strain $\boldsymbol{\varepsilon} = \boldsymbol{\varepsilon} + \mathbf{B} d\boldsymbol{\sigma}$ explicitly.

    \item \textbf{Iterate}:
    Repeat from step 1 until the loading is completed.
\end{enumerate}

\begin{figure}[H]
    \centering
    \includegraphics[width=0.85\linewidth]{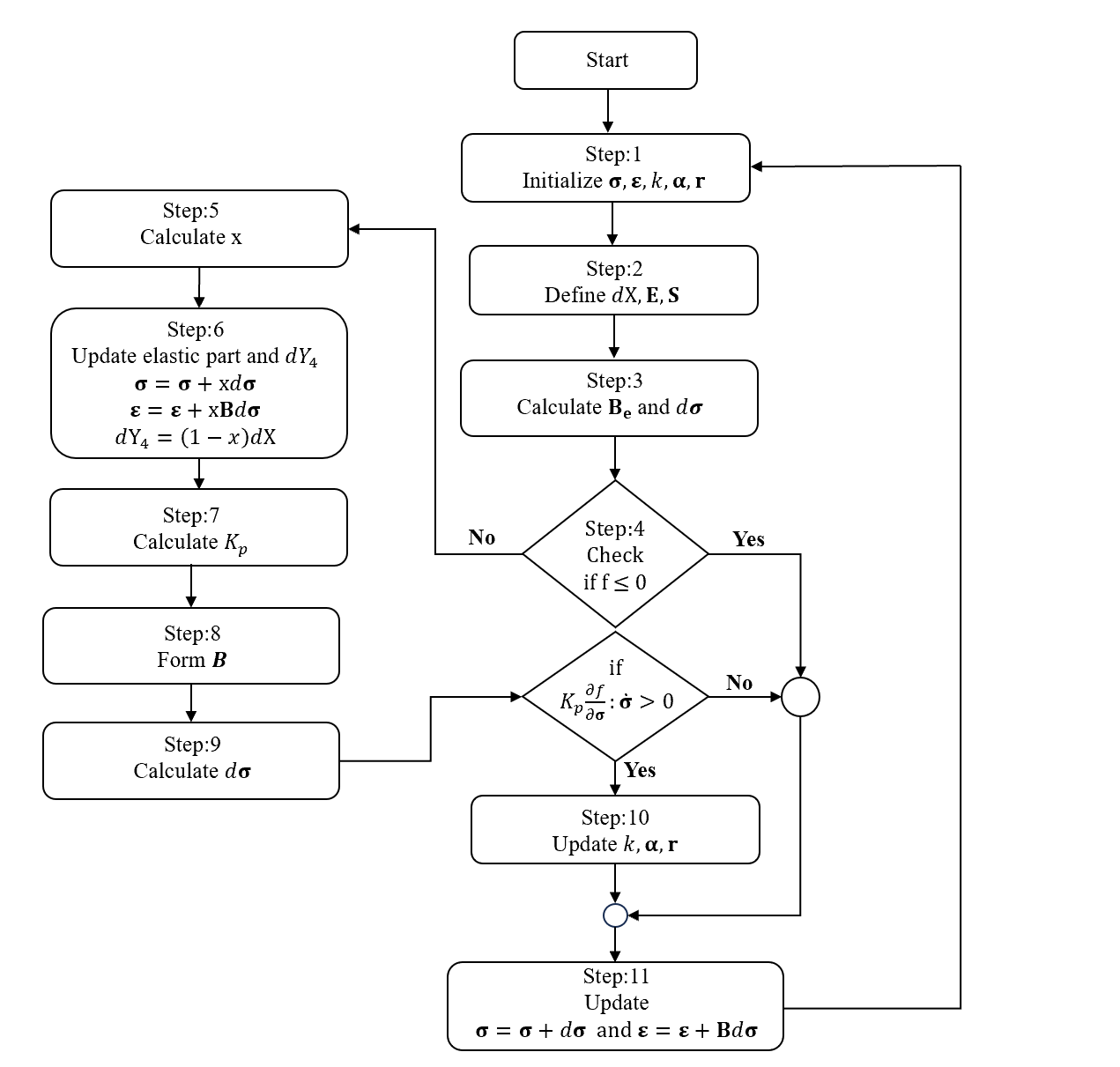}
    \caption{ Flowchart illustrating the numerical procedure by Bardet and Choucair \cite{bardet1991linearized} for incremental elastoplastic stress-strain analysis}
    \label{fig:fg3}
\end{figure}

The above structured approach can be shown using a flowchart in \cref{fig:fg3}
\clearpage

\bibliographystyle{ieeetr}
\bibliography{ref}

@article{wu1991experimental,
  title={On the experimental determination of yield surfaces and some results of annealed 304 stainless steel},
  author={Wu, Han C and Yeh, Wei C},
  journal={International Journal of Plasticity},
  volume={7},
  number={8},
  pages={803--826},
  year={1991},
  publisher={Elsevier}
}

@article{phillips1974some,
  title={Some new observations on yield surfaces},
  author={Phillips, A and Tang, J-L and Ricciuti, M},
  journal={Acta Mechanica},
  volume={20},
  number={1},
  pages={23--39},
  year={1974},
  publisher={Springer}
}

@book{mccomb1960some,
  title={Some experiments concerning subsequent yield surfaces in plasticity},
  author={McComb, Harvey G},
  year={1960},
  publisher={National Aeronautics and Space Administration}
}

@article{naghdi1958experimental,
journal = {Journal of Applied Mechanics},
  title={An experimental study of initial and subsequent yield surfaces in plasticity},
  author={Naghdi, Paul Mansour and Essenburg, F and Koff, W},
  year={1958},
volume = {25},
    number = {2},
    pages = {201-209},
    month = {06},

  publisher={American Society of Mechanical Engineers}
}

@article{boucher1995experimental,
  title={Experimental studies of yield surfaces of aluminium alloy and low carbon steel under complex biaxial loadings},
  author={Boucher, M and Cayla, P and Cordebois, J-P},
  journal={European journal of mechanics. A. Solids},
  volume={14},
  number={1},
  pages={1--17},
  year={1995}
}

@article{baltov1965rule,
  title={A rule of anisotropic hardening},
  author={Baltov, A and Sawczuk, A},
  journal={Acta Mechanica},
  volume={1},
  pages={81--92},
  year={1965},
  publisher={Springer}
}

@article{dafalias2003simple,
  title={A simple model for describing yield surface evolution during plastic flow},
  author={Dafalias, Yannis F and Schick, David and Tsakmakis, Charalampos},
  journal={Deformation and Failure in Metallic Materials},
  pages={169--201},
  year={2003},
  publisher={Springer}
}

@article{10.1115/1.2897056,
    author = {Voyiadjis, George Z. and Foroozesh, Mehrdad},
    title = {Anisotropic Distortional Yield Model},
    journal = {Journal of Applied Mechanics},
    volume = {57},
    number = {3},
    pages = {537-547},
    year = {1990},
    month = {09},
}

@article{ortiz1983distortional,
  title={Distortional hardening rules for metal plasticity},
  author={Ortiz, Miguel and Popov, Egor P},
  journal={Journal of Engineering Mechanics},
  volume={109},
  number={4},
  pages={1042--1057},
  year={1983},
  publisher={American Society of Civil Engineers}
}

@article{kurtyka1996evolution,
  title={Evolution equations for distortional plastic hardening},
  author={Kurtyka, T and {\.Z}yczkowski, M},
  journal={International Journal of Plasticity},
  volume={12},
  number={2},
  pages={191--213},
  year={1996},
  publisher={Elsevier}
}

@article{franccois2001plasticity,
  title={A plasticity model with yield surface distortion for non proportional loading},
  author={Fran{\c{c}}ois, Marc},
  journal={International Journal of Plasticity},
  volume={17},
  number={5},
  pages={703--717},
  year={2001},
  publisher={Elsevier}
}

@article{feigenbaum2007directional,
  title={Directional distortional hardening in metal plasticity within thermodynamics},
  author={Feigenbaum, Heidi P and Dafalias, Yannis F},
  journal={International Journal of Solids and Structures},
  volume={44},
  number={22-23},
  pages={7526--7542},
  year={2007},
  publisher={Elsevier}
}

@article{feigenbaum2008simple,
  title={Simple model for directional distortional hardening in metal plasticity within thermodynamics},
  author={Feigenbaum, Heidi P and Dafalias, Yannis F},
  journal={Journal of Engineering Mechanics},
  volume={134},
  number={9},
  pages={730--738},
  year={2008},
  publisher={American Society of Civil Engineers}
}

@article{ishikawa1997subsequent,
  title={Subsequent yield surface probed from its current center},
  author={Ishikawa, Hiromasa},
  journal={International Journal of Plasticity},
  volume={13},
  number={6-7},
  pages={533--549},
  year={1997},
  publisher={Elsevier}
}

@article{lemaitre1993mechanics,
  title={Mechanics of Solid Materials},
  author={Lemaitre, J and Chaboche, JL and Maji, Arup K},
  journal={Journal of Engineering Mechanics},
  volume={119},
  number={3},
  pages={642--643},
  year={1993},
  publisher={American Society of Civil Engineers}
}

@article{shutov2012viscoplasticity,
  title={A viscoplasticity model with an enhanced control of the yield surface distortion},
  author={Shutov, AV and Ihlemann, J},
  journal={International Journal of Plasticity},
  volume={39},
  pages={152--167},
  year={2012},
  publisher={Elsevier}
}

@article{panhans2006viscoplastic,
  title={A viscoplastic material model of overstress type with a non-quadratic yield function},
  author={Panhans, Sonja and Krei{\ss}ig, Reiner},
  journal={European Journal of Mechanics-A/Solids},
  volume={25},
  number={2},
  pages={283--298},
  year={2006},
  publisher={Elsevier}
}

@article{shutov2011phenomenological,
  title={A phenomenological model of finite strain viscoplasticity with distortional hardening},
  author={Shutov, AV and Panhans, S and Krei{\ss}ig, R},
  journal={ZAMM-Journal of Applied Mathematics and Mechanics/ Zeitschrift f{\"u}r Angewandte Mathematik und Mechanik},
  volume={91},
  number={8},
  pages={653--680},
  year={2011},
  publisher={Wiley Online Library}
}

@phdthesis{marek2019numerical,
  title={Numerical Implementation of Distortional Hardening Models},
  author={Marek, Ren{\'e}},
  year={2019},
  school={Czech Technical University}
}

@article{bardet1991linearized,
  title={A linearized integration technique for incremental constitutive equations},
  author={Bardet, JP and Choucair, W},
  journal={International Journal for Numerical and Analytical Methods in Geomechanics},
  volume={15},
  number={1},
  pages={1--19},
  year={1991},
  publisher={Wiley Online Library}
}

@article{janda2017general,
  title={General method for simulating laboratory tests with constitutive models for geomechanics},
  author={Janda, Tom{\'a}{\v{s}} and Ma{\v{s}}{\'\i}n, David},
  journal={International Journal for Numerical and Analytical Methods in Geomechanics},
  volume={41},
  number={2},
  pages={304--312},
  year={2017},
  publisher={Wiley Online Library}
}

@incollection{hutter2018coleman,
  author    = {Geralf Hütter},
  title     = {Coleman--{N}oll Procedure for Classical and Generalized Continuum Theories},
  booktitle = {Encyclopedia of Continuum Mechanics},
  editor    = {Holm Altenbach and Andreas Öchsner},
  publisher = {Springer},
  address   = {Berlin, Heidelberg},
  year      = {2018},
  doi       = {10.1007/978-3-662-53605-6_57-1}
}

\end{document}